\documentclass[a4paper, conference]{IEEEtran}
\usepackage{cite}
\usepackage[pdftex]{graphicx}

\begin{document}

\title{Performance Evaluation of Parallel TCP, and its Impact on Bandwidth Utilization and Fairness in  High-BDP Networks Based on Test-bed}


\author{\authorblockN{Mohamed A. Alrshah, Mohamed Othman}
\authorblockA{Department of Communication Technology and Networks\\
Universiti Putra Malaysia\\
43400, UPM, Serdang, Selangor D.E, Malaysia.}
mohamed.asnd@gmail.com, mothman@upm.edu.my}

\maketitle


\begin{abstract}
After the presence of high Bandwidth-Delay Product (high-BDP) networks, many researches have been conducted to prove either the existing TCP variants can achieve an excellent performance without wasting the bandwidth of these networks or not. In this paper, a comparative test-bed experiment on a set of high speed TCP variants has been conducted to show their differences in bandwidth utilization, loss ratio and TCP-Fairness. The involved TCP Variants in this experiment are: NewReno, STCP, HS-TCP, H-TCP and CUBIC. These TCP variants have been examined in both cases of single and parallel schemes. The core of this work is how to evaluate these TCP variants over a single bottleneck network using a new parallel scheme to fully utilize the bandwidth of this network, and to show the impact of accelerating these variants on bandwidth utilization, loss-ratio and fairness. The results of this work reveal that, first: the proposed parallel scheme strongly outperforms the single based TCP in terms of bandwidth utilization and fairness. Second: CUBIC achieved better performance than NewReno, STCP, \mbox{H-TCP} and HS-TCP in both cases of single and parallel schemes. Briefly, parallel TCP scheme increases the utilization of network resources, and it is relatively good in fairness.
\end{abstract}

\vspace{0.25cm}



\section{Introduction}


The ability to quickly move a large amounts of data through a shared network is very important for several applications today. Some of these applications need to move thousands of gigabytes of data yearly between computers that are very far from each other. Multi-users collaborative environments, that uses visualization, video, voice, and remote desktop, require low network latency and high throughput. The Optiputer project aims to build a distributed high performance computer using global optical networks as the system backplane. likewise, PSockets, GridFTP, DPSS and BBCP need to move a large amounts of data through a shared network \cite{hacker2004b}.

Many of these applications deploy the Transmission Control Protocol (TCP) for accuracy and reliability in order to transmit data. TCP relies on the congestion avoidance scheme to: (1) estimate the bandwidth of the network route, (2) share this estimated bandwidth fairly between the competing TCP flows, and (3) maximize the efficiency on bandwidth utilization.

On a shared network, the rewards for aggressive behavior are not balanced with penalties for misbehavior that would encourage the fair sharing of network bandwidth. This creates a tragedy of the common situations, in which application's net gain results in a net loss borne by the community of users that choose to act cooperatively. Thus, the problem of providing mechanisms for reliable high throughput transmission on shared networks, that can overcome the limitations of TCP congestion avoidance and fairly share limited network resources, is an important problem that needs to be solved.

\subsection{Relationship between Packet Loss and TCP Performance}

The implication of the relationship between packet loss and TCP performance is that, if the source of any packet loss is not due to network congestion, the number of non-congestion losses over a period of time could be large enough to adversely affect TCP performance. The Mathis \cite{Mathis1997} and Padhye \cite{padhye1998} TCP bandwidth estimation equations state that, TCP throughput is inversely proportional to the square root of the packet loss rate. Because of this relationship, TCP performance over high-BDP networks requires incredibly low non-congestion packet loss rates for the congestion avoidance algorithm to successfully probe network capacity. For example, using the Mathis equation \cite{Mathis1997}, the packet loss rate must be less than or equal to $0.0018\%$ or $2/100000$ packets to allow a TCP flow to utilize at least 2/3 of a 622 Mbps Optical Carrier (OC-12) ATM link. Floyd found that the maximum permitted IEEE bit error rate (BER) for a fiber optic line is large enough to prevent a TCP flow from ever making full use of a 10 Gbps Ethernet network over a transoceanic link \cite{hacker2004}.

\subsection{Approaches to Solve TCP Performance Problems}

An approach commonly used to solve TCP performance problems is to create multiple TCP flows to simultaneously transmit data over several sockets between an application server and client. PSockets \cite{sivakumar2000} provides an application library which can be used by an application to stripe data transmissions over a set of parallel TCP flows. The use of parallel TCP flows has also been adopted by GridFTP \cite{allcock2005}, MulTCP \cite{crowcroft1998}, BBCP \cite{hanushevsky2001}, DPSS \cite{Tierney1994}, and other high performance data intensive applications. Parallel TCP is an aggressive approach that can overcome the effects of non-congestion loss, but it does so at the expense of unfairly appropriating bandwidth from competing TCP flows when there is a limited network capacity. Other approaches, to improve performance, have been proposed, but all of them suffer from the same problem: effectiveness is increased, but at the expense of fairness \cite{hacker2004}. In fact, it is feasible and valuable to build a new network protocol, based on parallel TCP scheme for data intensive applications, to fully utilize the bandwidth of high-BDP networks and maintains fairness.

\section{Literature Review}

To increase the utilization of network bandwidth, many TCP variants have been developed. Scalable TCP (STCP) \cite{kelly2003} is a simple change to the traditional TCP \cite{RFC2581} which significantly enhances the bandwidth utilization of TCP over high-BDP networks. As known, the traditional TCP relies on the transmission rate and round trip time. Unlikely, STCP relies only on the round trip time which increases the scalability especially over high-BDP networks \cite{bateman2008}.

High speed TCP (HS-TCP) \cite{RFC3649} modifies the standard TCP scheme to conquer its limitations. In congestion avoidance, HS-TCP increases its congestion window by $a(w)/w$ after every reception of ACK,  while it decreases the congestion window by $(1 - b(w)) w$ after loss detection, where $w$ is the last congestion window registered before the last loss detection. Indeed, HS-TCP works similar to the standard TCP when the \textit{cwnd} is small but if TCP's \textit{cwnd} is beyond a certain threshold, it increases by $a(w)$ and decreases by $b(w)$ that are functions of the last congestion window registered before the last loss detection. Thus, the growth of HS-TCP \textit{cwnd} is faster than the standard TCP which makes the loss recovery more faster and increases the bandwidth utilization in high-BDP networks as well \cite{bateman2008}.

In addition, H-TCP \cite{shorten2004} increases the bandwidth utilization of TCP on high-BDP networks, and also maintains the fairness \cite{bateman2008}. As the time since the last loss detection increases, H-TCP increases its \textit{cwnd}. This eliminates the problem of HS-TCP \cite{RFC3649} and BIC \cite{xu2004} which increases the aggressiveness of the flow if its \textit{cwnd} is already large. Thus, the new established flows converges to fairness more faster under H-TCP than HS-TCP and BIC. In some cases, the behavior of H-TCP can lead to unfair share of network bandwidth \cite{bateman2008}. Furthermore, CUBIC \cite{ha2008} is a derivative version of BIC designed for high-BDP networks. It succeeded to increase the bandwidth utilization and fairness especially in high-BDP networks \cite{bateman2008}.

Alternatively, the parallel TCP approach has been used to overcome the limitations of single based TCP on bandwidth utilization over high-BDP networks \cite{lekashman1989, iannucci1992, allman1996}. The concept of Parallel TCP is the use of a set of multiple standard or modified TCP flows. Recently, many applications, such as GridFTP \cite{allcock2002,allcock2005}, PSockets \cite{sivakumar2000,grossman2003} and QTCP \cite{qureshi2012}, have been developed to improve the bandwidth utilization over high-BDP networks by using multiple TCP flows. Additionally, some solutions are applying the aggregated congestion window of the logical parallel flows, as proposed in \cite{balakrishnan1999,eggert2000,hacker2004b,hacker2002}, which reduces the effect of packet loss and increases the TCP's bandwidth utilization.

Unlikely, some TCP variants have the ability to act like parallel, emulated by using single based TCP such as MulTCP, which makes one connection acts like a set of concurrent TCP flows in order to achieve weighted proportional fairness. It has been said that, the parallel TCP can be emulated and well replaced by a single based TCP which properly modified to act like parallel based TCP. Otherwise, the existing single based TCP will not be able to fully utilize the high speed bandwidths which provided by the high-BDP networks \cite{fu2007}.

However, the main problem of using a single based TCP, emulating a set of multiple standard or modified TCP flows, is when an ack received, the aggregated window size will increase by a certain number of packets based on the TCP variant which is used, and it can quickly grow but when timeout detected, the aggregated window size will be decreased to the half of the previous window which will affect the bandwidth utilization. While in parallel of multiple standard or modified TCP flows, each flow works separately from the concurrent flows, when one of them detects a timeout it will only decreases the window size of the involved flow. While, the other concurrent flows will keep their window increase until the timeouts detected. This approach highly improves the bandwidth utilization and makes the TCP flows behave fairly with each other as in the single TCP approach.

Moreover, single based TCP has been designed to use a single path between the source and destination. Thus, multi-paths cannot be used in the single based approach. To support multi-paths use, parallel TCP (pTCP) has been developed. pTCP allows connections to use the aggregated bandwidths of the multiple paths, regardless of the characteristics of every path \cite{hsieh2002}.

From the observation on parallel TCP, it has been found that, the parallel TCP is more effective than the single based TCP especially in high-BDP networks. Single based approach achieves lower bandwidth utilization than the parallel TCP \cite{fu2007}. But on shared network bottleneck, the parallel TCP unfairly steals bandwidth from the competing single based TCP flows. To improve the total performance of TCP, bandwidth utilization should be improved while maintaining fairness \cite{hacker2004b}.

\section{Fast Recovery in Single Based TCP}

\begin{figure}[h]
\centerline{\includegraphics[width=7cm]{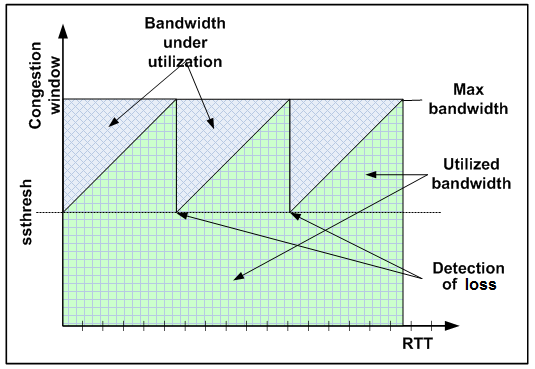}}
\caption{Evolution of congestion window in single based TCP.}
\label{evo-single}
\end{figure}

Assume that, the periodic loss event has been used; the evolution of \textit{cwnd} for a single connection when Fast Recovery is taken into account will be as shown in Figure \ref{evo-single}. Which means that, after a timeout detected; AIMD will halve its \textit{cwnd}. This will affect the whole throughput of the connection and it will take a long time to reach the maximum \textit{cwnd} again. As shown in Figure \ref{evo-single}, the green colored area reflects the throughput of the connection while the gray colored area reflects the unutilized area. It is very clear that, detection of one timeout signal can reduce the \textit{cwnd} to approximately $50\%$. This considered as a problem of wasting resources, this problem has been partially solved in parallel TCP and it will be explained in the next section.

\section{Fast Recovery in Parallel TCP}

\begin{figure}[h]
\centerline{\includegraphics[width=7cm]{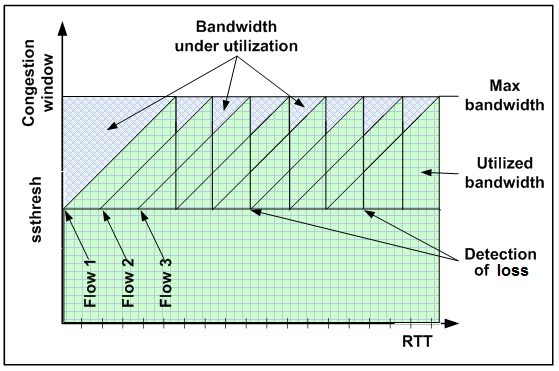}}
\caption{Evolution of congestion window in Parallel TCP.}
\label{evo-parallel}
\end{figure}

With the same assumptions in the previous section, Figure \ref{evo-parallel} shows the evolution of the \textit{cwnd} for three concurrent flows belongs to one parallel scheme. It is well clear that, the detection of timeout in one connection will decrease the \textit{cwnd} of the involved flow only while the other concurrent flows will not be affected and they will continue in their \textit{cwnd} increasing until the timeouts detected. The main two reasons to make parallel TCP behave in this way are the serialization of flows establishments, and the independence of concurrent flows.

Assume that, the available link bandwidth was only one Mbps, and there are three flows share this link; also assume that, the bandwidth was equally divided between the flows, which mean approximately 0.33 Mbps for each flow. If a timeout detected on Flow 1 it will decrease its \textit{cwnd} to the half, which is around 0.16 Mbps, while the \textit{cwnds} of the others will stay as they are (0.33 Mbps for each). The aggregation window for these three concurrent flows after packet loss detection will be as shown in Equation (\ref{eq1}) below,

\begin{equation}
\label{eq1}
ACW = CWflow1 + CWflow2 + CWflow3
\end{equation}
Thus, $ACW \simeq 0.16 + 0.33 + 0.33  \simeq 0.82 Mbps$.
\newline\newline
While, \emph{ACW} is the aggregated \textit{cwnd} of the parallel flows and \emph{CW} is \textit{cwnd} of a single flow. Consequently, the reduction of the \textit{cwnd} in parallel of three flows will be as shown in Equation~(\ref{eq2}).

\begin{equation}
\label{eq2}
Reduction = \frac{Max ACW - CurrentACW}{Max ACW}
\end{equation}

Thus, $Reduction \simeq \frac{1 Mbps - 0.82 Mbps}{1 Mbps} \simeq 0.18\%$
\vspace{0.5cm}

\section{Multi-Route Concept}

In single based TCP, the connection does not have the ability to use more than one route at a single time. During the connection establishment the intermediate routers will chose one route to be used by this connection but if route failure has been detected after a certain amount of time, the intermediate routers will change to another route from the available routes. This will ensure the use of single route, this considered as a problem of wasting resources, especially when multi-paths infrastructure is available, because in some scenarios the chosen route limits the connection capability while there are an alternative routes that are free or not fully utilized.

\begin{figure}[h]
\centerline{\includegraphics[width=\linewidth]{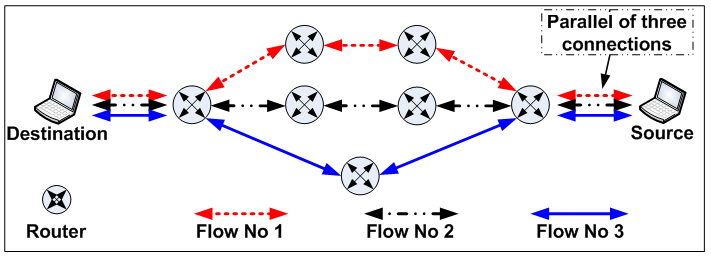}}
\caption{Multi-Route Usage.}
\label{multi-route}
\end{figure}

Contrarily, the proposed parallel TCP can utilize multi-routes without any modification this resulted by the independence of the parallel flows. Assume that, there are three parallel flows belongs to one application process, during the connections establishments, the application will start with the first connection and the intermediate routers will chose one of the available routes to be used by this connection. Then the application will establish the next connection, and the intermediate routers may chose the same route, which already used by the first connection, or another route from the available routes as shown in Figure \ref{multi-route}. This selection of route relies on some criteria such as the link utilization, distance, link delay and link cost. The use of multiple routes will increase the utilization of the available resources and thus will increase the throughput of these parallel TCP flows.

\section{Research Methodology}

The goal of this work is, to develop a new parallel TCP algorithm, using a set of TCP sessions for data intensive applications, in order to solve the performance problems of single based TCP and to effectively utilize the high-BDP network links while maintaining fairness. This algorithm has been built in $C\#$ using MonoDevelop IDE over Linux openSuse 12.2. For evaluation purpose, a new test-bed experiment scheme \cite{AlrshahThesis} has been developed and used. This experiment aims to show the impact of using the proposed algorithm on bandwidth utilization, loss ratio and TCP fairness. 

\begin{figure}[h]
\centerline{\includegraphics[scale=0.3]{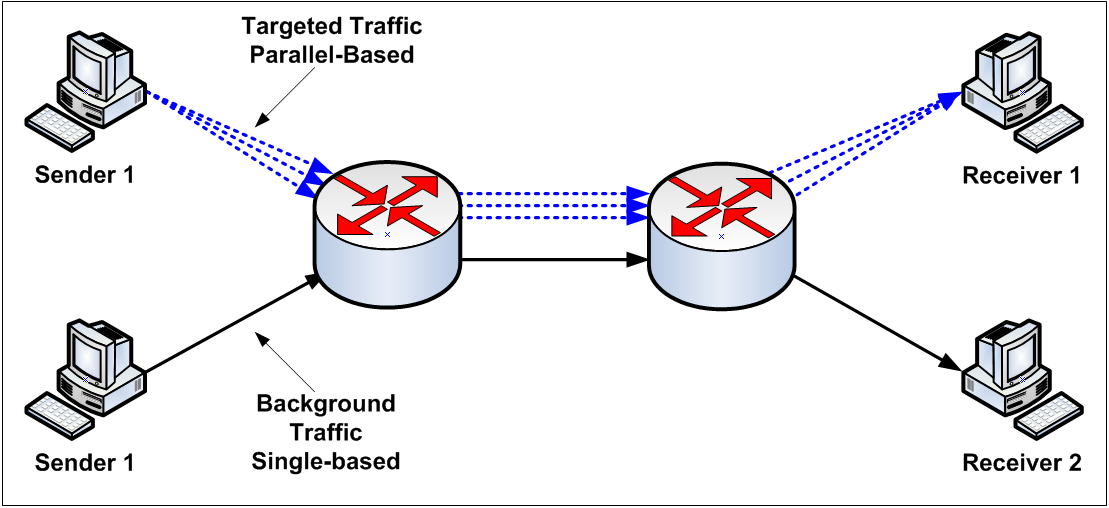}} 
\caption{Network Topology.}
\label{topo}
\end{figure}

The proposed algorithm has been examined using a number of high speed TCP variants that are NewReno, Scalable, H-TCP, HS-TCP and CUBIC. This experiment has been done using a single-bottleneck topology, over high-BDP wired network. Figure \ref{topo} \cite{alrshah2009} shows the network topology which is a typical single dumbbell topology, with a symmetric channel and loss-free reverse path while Table \ref{params} shows the experiment setup. The targeted traffic has been observed in presence of the background traffic. The performance metrics, that are bandwidth utilization, loss ratio and fairness, have been observed. In addition, Jain's Fairness Index (JFI) \cite{chiu1989,jain1984} has been used in this paper as shown in Equation \ref{eq3} to measure the TCP fairness.

\begin{equation}
\label{eq3}
TCP Fairness (F) = \frac{(\sum^{N}_{i=1}
X_{i})^{2}}{N(\sum^{N}_{i=1} X^{2}_{i})}
\end{equation}

Where $X_i$ is the measured throughput for $flow_i$, from $N flows$ in the system.

\begin{table}[h]
	\caption{Experiment Parameters}
	\begin{center}
	\begin{tabular}{p{0.2cm}p{3cm}p{4.2cm}} 
	\hline
	No.& Parameter				 &	Value										\\ \hline
	1. & TCP Scheme			 	 &	newReno/Scalable/Htcp/HStcp/CUBIC			\\ 
	2. & Flow Control Algorithm  &	Random Early Drop (RED)						\\ 
	3. & Link capacity			 &	100 Mbps for nodes and 10 Mbps for			\\ 
	   & 						 &	bottleneck									\\ 
	4. & Link delay			 	 &	100 milliseconds							\\ 
	5. & Bandwidth Delay Product &	125000 Bytes (High-BDP as in \cite{RFC1072})\\ 
	6. & Packet size			 &	1000 bytes									\\ 
	7. & Buffer size			 &	300 packets									\\ 
	8. & Traffic type			 &	Standard Poisson distribution				\\ 
	9. & Experiment time		 &	1000 seconds								\\ \hline
	\end{tabular}
	\label{params}
	\end{center}
\end{table}

\section{Results and Discussion}

\begin{figure}[h]
\centerline{\includegraphics[width=8cm]{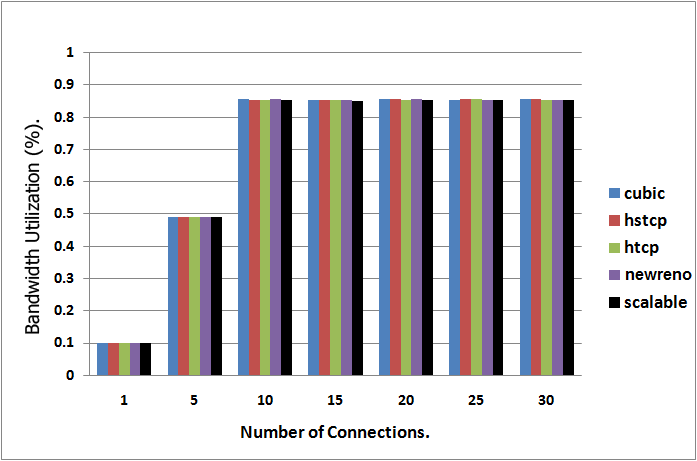}} 
\caption{bandwidth utilization vs. Number of Connections.}
\label{throughput}
\end{figure}

The results of this experiment as shown in Figure \ref{throughput} reveal that, the bandwidth utilization of the involved TCP variants is almost the same in all cases, which means that, all of these TCP variants can achieve similar bandwidth utilization and can provide the same link utilization. Moreover, all of these TCP variants achieve higher performance in parallel modes than single mode. Also the bandwidth of the bottleneck link has not been fully utilized in both cases of single flow and parallel of 5 TCP flows, while it is almost fully utilized in the rest cases (parallel of 10, 15, 20, 25, 30 TCP flows).

\begin{figure}[h]
\centerline{\includegraphics[width=8cm]{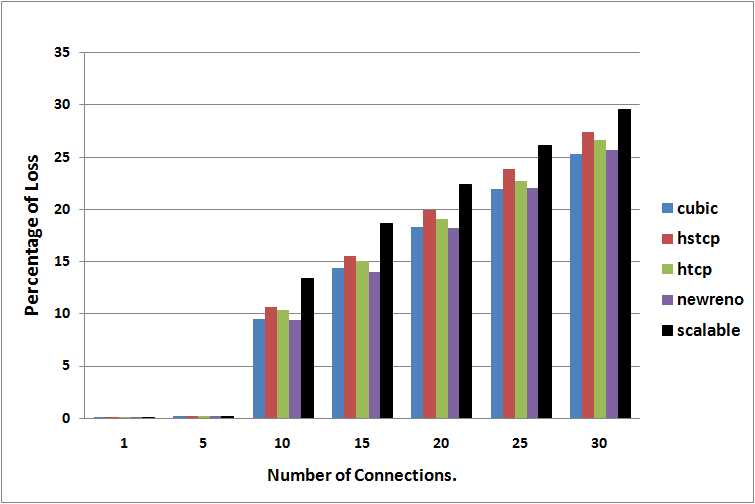}} 
\caption{Loss Ratio vs. Number of Connections.}
\label{loss}
\end{figure}

\begin{figure}[h]
\centerline{\includegraphics[width=8cm]{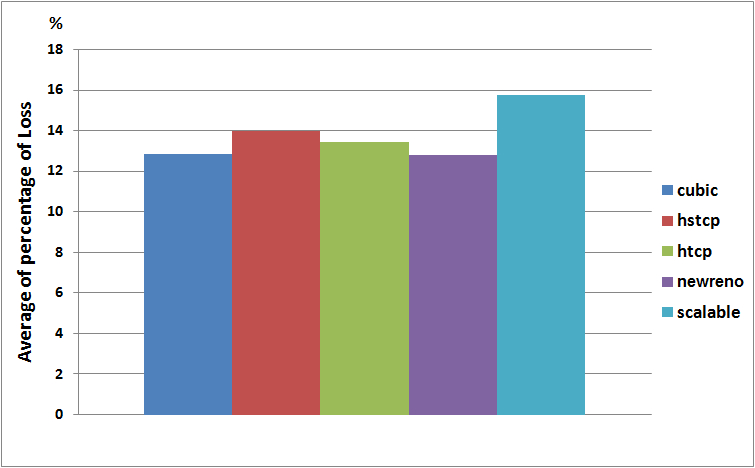}} 
\caption{The average of loss ratio among TCP variants.}
\label{avg-loss}
\end{figure}

On the other hand, Figure \ref{loss} shows the main difference between TCP variants which is the loss ratio. The graph reveals that, the use of Scalable, H-TCP and HS-TCP cause a lot of unnecessary retransmission while CUBIC and NewReno do not. Due to the conservative \textit{cwnd} increase, CUBIC and NewReno achieved better performance than the other TCP variants. Figure \ref{avg-loss} shows the average of loss ratio among TCP variants, this graph gives a brief comparison to facilitate the operation of TCP evaluation, and it is so clear that, CUBIC and NewReno were the best while TCP Scalable was the worst.

As shown in Figure \ref{loss}, when the bottleneck is not fully utilized (no congestion), in both cases of single TCP connection and parallel of 5 TCP flows, the loss ratio is very small. But, after the number of parallel TCP flows increased the loss ratio increases as well. In Figure \ref{throughput}, the bottleneck link of the implemented topology has been fully utilized when the number of parallel TCP flows equals to 10 TCP flows. This means that, the increasing of the number of parallel TCP flows more than 10 flows for this link capacity will not be useful and it will cause TCP overhead which is the increasing of unnecessary retransmission while the bandwidth utilization cannot exceed the bandwidth limit of the bottleneck link.

Figure \ref{avg-loss} shows that, the averages of loss ratio that recorded by these TCP variants were clearly different, for instance, Scalable TCP was the worst one and it has recorded around $16\%$ of data loss from the entire throughput, contrarily CUBIC and NewReno was the best ones and they have recorded around $13\%$ of data loss from the entire throughput. Moreover, H-TCP and HS-TCP are partially reasonable and better than Scalable TCP. 

\begin{figure}[h]
\centerline{\includegraphics[width=7.9cm]{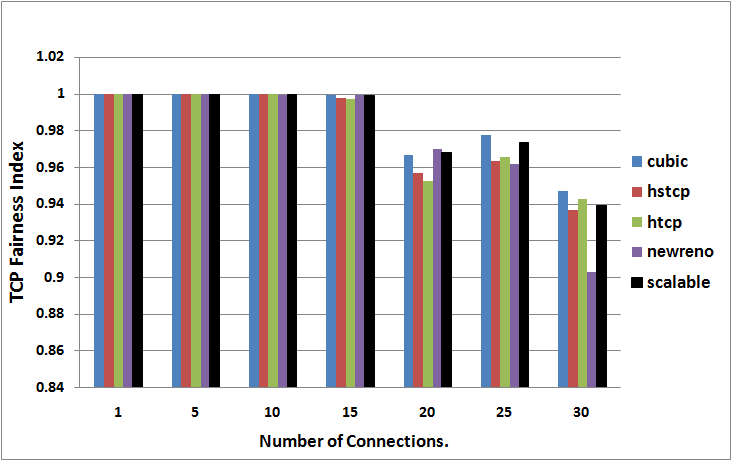}}
\caption{TCP Fairness Index vs. Number of Connections.}
\label{fairness-index}
\end{figure}

\begin{figure}[h]
\centerline{\includegraphics[width=7.9cm]{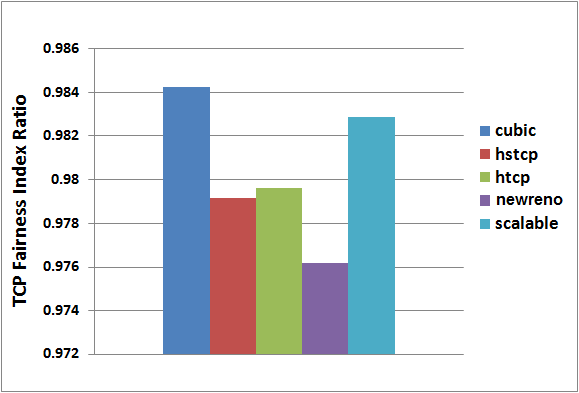}}
\caption{TCP Fairness Index Ratio among TCP variants.}
\label{fairness-ratio}
\end{figure}

Relatively, the range of fairness variation starts from $100\%$ to $90\%$. As shown in Figure \ref{fairness-index}, NewReno scores the worst fairness index compared with the others while CUBIC and Scalable were the best, but Scalable was highly affected by increasing the number of flows, unlikely, CUBIC was reasonably affected. When the bottleneck is not congested, all of the TCP variants were almost the same and they achieve similar Fairness Index which is about $100\%$ but after increasing the number of TCP flows to be more than 10 flows, which makes the bottleneck highly congested, the fairness index is slightly decreased. Figure \ref{fairness-ratio} shows in brief that, the order of TCP variants based on TCP Fairness is CUBIC, Scalable, H-TCP, HS-TCP and NewReno. The first order the highest fairness and the last order the worst Fairness.

Clearly, for this topology with 10 Mbps bottleneck, the appropriate number of parallel TCP flows, to fully utilize the bandwidth, is 10 concurrent TCP flows, that will cause the least possible amount of unnecessary retransmissions with high fairness index. This number of parallel TCP flows considered as the threshold of TCP parallelism of this bottleneck bandwidth, and each link capacity has a proper parallelism threshold. This threshold should be carefully calculated before starting TCP parallelism based on some variables that are not in the scope of this work.

\section{Conclusion}

It is clearly concluded that, (1) single based TCP cannot overcome parallel TCP especially in high-BDP networks, (2) all TCP variants, in parallel mode, can achieve a good bandwidth utilization but when the bottleneck is fully utilized, which means that, there is a congestion, the difference will not be in the bandwidth utilization but it will be in loss ratio and TCP fairness, (3) CUBIC TCP achieves higher performance than the other TCP variants in terms of bandwidth utilization, loss-ratio and fairness, (4) The proposed parallel TCP algorithm achieves high bandwidth utilization and it can effectively utilize high-BDP network links while it maintains the fairness among the competed flows.

In this experiment, some of TCP features like SACK and FACK have been disabled to show the impact of changing the congestion control algorithms on TCP performance, but there is a strong intention to repeat this experiment with different settings. For instance, SACK and FACK may be enabled to emphasize their impact and to show either they are worth to be used with the proposed parallel TCP algorithm or not. On the other hand, there are some modifications have to be done later in Linux kernel to implement the proposed parallel TCP.

\medskip{\bf Acknowledgment}
This work was supported by the Ministry of Higher Education of Malaysia under the Fundamental Research Grant FRGS/02/01/12/1143/FR for financial support.

\begingroup
\let\itshape\upshape


\endgroup



\end{document}